\documentclass[reprint,aps,pre,superscriptaddress]{revtex4-1}

\usepackage{textcomp}
\usepackage{amsmath}
\usepackage{amssymb}
\usepackage{mathrsfs}
\usepackage[retainorgcmds]{IEEEtrantools}
\usepackage[normalem]{ulem}
\usepackage{color}
\usepackage{empheq}
 
\usepackage{graphicx}

\begin{document}


\title{Magnetic phase transition in coupled spin--lattice systems:\\ A replica-exchange Wang--Landau study}
%


\author{Dilina Perera}
\email[]{dilinanp@physast.uga.edu}
\affiliation{Center for Simulational Physics, The University of Georgia, GA 30602, USA}
\affiliation{Department of Physics and Astronomy, Mississippi State University, Mississippi State, Mississippi 39762, USA}

\author{Thomas Vogel}
\affiliation{Department of Physics, Stetson University, DeLand, FL 32723, USA}

\author{David P. Landau}
\affiliation{Center for Simulational Physics, The University of Georgia, GA 30602, USA}



\begin{abstract}
  Coupled, dynamical spin--lattice models provide a unique
  test ground for simulations investigating the
  finite-temperature magnetic properties of materials under
  the direct influence of the lattice vibrations.  These models are
  constructed by combining a coordinate-dependent interatomic
  potential with a Heisenberg-like spin Hamiltonian, facilitating the
  treatment of both the atomic coordinates and spins as explicit phase
  variables.  Using a model parameterized for bcc iron, we
  study the magnetic phase transition in these complex
  systems via the recently introduced, massively parallel
  replica-exchange Wang--Landau Monte Carlo method.  Comparison with
  the results obtained from rigid lattice (spin only) simulations show
  that the transition temperature as well as the amplitude of the peak
  in the specific heat curve is marginally affected by the lattice
  vibrations.  Moreover, the results were found to be sensitive to the
  particular choice of the interatomic potential.
\end{abstract}


\maketitle


\section{Introduction}

With the continuing developments in materials science and engineering,
a renewed interest has emerged in under\-standing the
temperature-dependent magnetic properties pertaining to real
materials.  This demands sophisticated and improved magnetic models
that are capable of providing a more realistic depiction of the
material than that is possible with conventional spin models.  A~novel
class of such improved models that continues to gain widespread
attention are atomistic models that treat the dynamics of the
translational (atomic) degrees of freedom on an equal footing with the
spin (magnetic) degrees of freedom~\cite{Ma2008, AnisCobalt,
  Omelyan2001, Thermodynamics, Omelyan2001}.  We will refer to such
models as (coupled, dynamical) spin--lattice models.  The motivation
for these hybrid models is the substantial amount of experimental and
theoretical evidence that suggests strong phonon--magnon coupling
in magnetic crystals, particularly in transition metals and
alloys~\cite{SpinLatticeCoupling1, SpinLatticeCoupling2}.
A~parameterized spin--lattice model for bcc iron developed by Ma
\textit{et~al.}~\cite{Ma2008} has been subjected to a number of
subsequent studies targeted towards understanding the dynamical
behavior, including vacancy formation and migration~\cite{WenVacancy1,
  WenVacancy2}, and phonon--magnon interactions~\cite{Perera2014,
  PereraST2014}.
Moreover, the model has been recently extended by incorporating spin-orbit interactions~\cite{spin_orbit}, which, 
in particular, extends its applicability to accurate modeling of non-equilibrium dynamical processes.

Previous work on coupled spin--lattice systems was almost exclusively
performed using the combined molecular and spin dynamics
technique~\cite{Ma2008, PereraST2014}, in which the coupled equations
of motion for all degrees of freedom are simultaneously solved to
obtain phase-space trajectories in real time.  A single study has been
reported where parallel tempering Monte Carlo (MC) method was applied
to relatively small system sizes to investigate the magnetic phase
transition in iron~\cite{Thermodynamics}.  In addition to the
  obvious inflation of the phase space due to the inclusion of the
  extra spatial degrees of freedom, the coupling between the spin and
  lattice subsystems may also pose a significant challenge for the
  sampling due to the emergence of novel excitations such as coupled
  phonon--magnon modes~\cite{Perera2014}.  Thus, the study of
reasonably large systems without compromising the accuracy and
efficiency requires state-of-the-art MC methods that effectively
utilize modern computing resources.

Among numerous MC methods introduced in the past few decades,
Wang--Landau sampling~\cite{wl1,wl2,wl3}
stands out as a powerful, yet a simple technique
with only a few adjustable parameters.  Unlike canonical MC methods in
which the goal is to generate a sequence of microstates from the
canonical ensemble at a given temperature $T$, the Wang--Landau method
strives to deliver an estimate of the density of states $g(E)$, where
$E$ is the energy, as the end product.  In essence, this is
accomplished by, ideally, performing a random walk in energy space
while iteratively adjusting the density of states.  The estimated
density of states can then be used to extract thermodynamic properties
for the entire temperature range of interest.  An inherent advantage
of Wang--Landau sampling is its ability to easily overcome free energy
barriers.  Thus the method has been frequently applied for systems
with rough free energy landscapes such as spin glasses, liquid
crystals, polymers and proteins etc.~\cite{spinGlass, liquidCrystal,
  polymer, proteinFolding}.

The recently introduced replica-exchange Wang--Landau (REWL)
framework~\cite{rewl1, rewl2, rewl3, rewl4, rewl5}
further pushes the limits of
Wang--Landau sampling by directly exploiting the power of the modern
parallel computing systems.  In this approach, the total energy range
is divided into a set of overlapping windows that are concurrently
sampled by independent random walkers.  Adopting the concept of
conformational swapping from parallel tempering~\cite{partemp1,
  partemp2}, occasional configurational (replica) exchanges between
overlapping windows are allowed, facilitating each replica to traverse
through the entire energy range.

In this paper, we explore the feasibility and the efficacy of using
the REWL method for coupled spin--lattice
systems that are specifically parameterized for bcc iron.  In
Sec.~\ref{sec:model_and_methods}, we describe the system Hamiltonian
and the parameterization that we adopt, and provide a detailed
description of the REWL method.  In
Sec.~\ref{sec:results}, we present our results and analysis, with
emphasis on exploring the impact of the phonons on the magnetic phase
transition, as well as the sensitivity of the results to different
interatomic potentials.

\section{Model and methods}~\label{sec:model_and_methods}

\subsection{Coupled spin--lattice Hamiltonian for bcc
  iron}~\label{sec:model}

Let us consider a classical system of $N$ magnetic atoms of mass $m$,
described by their positions $\{\mathbf{r}_i\}$ and the orientations
$\{\mathbf{e}_i\}$ of the atomic spins. The corresponding Hamiltonian
can be written as
\begin{equation} \label{eq:mdsd_hamiltonian}
  \mathcal{H} = U(\{\mathbf{r}_i\}) - \sum_{i<j} J_{ij}(\left\{ \mathbf{r}_k \right\}) \mathbf{e}_i \cdot \mathbf{e}_j,
\end{equation}
where $U(\{\mathbf{r}_i\})$ represents the spin-independent
(non-magnetic) scalar interaction between the atoms, and the
Heisenberg-like interaction with the coordinate-dependent exchange
parameter $J_{ij}(\left\{ \mathbf{r}_k \right\})$ specifies the
exchange coupling between the spins.  

Since the theoretical framework for interaction potentials that
specifically exclude magnetic contributions is not yet available,
we construct $U(\{\mathbf{r}_i\})$ as
\begin{equation} \label{eq:U_non_mag}
	U(\{\mathbf{r}_i\}) = U_{\text{EAM}}(\{\mathbf{r}_i\}) - E_{\text{spin}}^{\text{ground}},
\end{equation}
where $U_{\text{EAM}}$ represents a conventional interatomic potential
for bcc iron based on the embedded atom model (EAM), and
$E_{\text{spin}}^{\text{ground}} = - \sum_{i<j} J_{ij}(\left\{
  \mathbf{r}_k \right\})$ is the energy contribution from a collinear
spin state which we subtract to eliminate the magnetic interaction
energy implicitly contained in $U_\text{EAM}$.  With the chosen form
of $U(\{\mathbf{r}_i\})$, the Hamiltonian~\eqref{eq:mdsd_hamiltonian}
provides the same energy as $U_\text{EAM}$ for the ferromagnetic
ground state at $0$\;K.  

For $U_\text{EAM}$, we choose two well-established EAM potentials for
bcc iron, namely, the Finnis--Sinclair
potential~\cite{Finnis1984,Finnis1986}, and the Dudarev--Derlet
``magnetic'' potential~\cite{Dudarev2005,Derlet2007}.
Introduced in 1984, the Finnis--Sinclair (FS) model is one of
  the oldest and most frequently used many-body potentials for bcc
  iron.  The theoretical foundation of the FS potential is based on a
  second-moment approximation to the tight binding density of states.
  Despite its simple empirical form and the short cut-off distance,
  the FS potential can reproduce bulk material properties, such as
  bulk moduli and elastic constants, reasonably
  accurately~\cite{AboutFS1}. Hence, it has long been a popular choice
  among materials scientists. However, it is not suitable for modeling
  highly disordered systems such as interstitial and vacancy
  configurations since the repulsive part of the potential is too
  ``soft,'' and thus tends to produce nonphysical results for such
  systems~\cite{WhatsWrongFS1, WhatsWrongFS2}.

Among various empirical potentials derived for bcc iron, the
  recently introduced Dudarev--Derlet (DD) potential stands out due to
  its unique feature of taking the local magnetic structure into
  account when determining the interatomic forces. The DD potential is
  based upon the Stoner and the Ginzburg--Landau models and is motivated
  by the fact that the presence of magnetism significantly contributes
  to the stability of the crystal structure in iron-based
  materials~\cite{PhaseStabilityIron, PhaseStabilityFeCo}. It was then
  parameterized using a wide range of material properties, including
  bulk cohesive energy, lattice constants, elastic constants, and
  vacancy formation energies corresponding to both bcc and fcc
  configurations, as well as magnetic and non-magnetic
  phases~\cite{Dudarev2005}. The DD potential does not
  treat the orientational dynamics of the atomic moments, and
  therefore, the treatment of non-collinear spin configurations at
  finite temperatures is outside its domain of applicability. To
  achieve this, one needs to incorporate the dynamics of the spin
  orientations explicitly~\cite{Ma2008}.

For modeling the exchange interaction $J_{ij}(\left\{ \mathbf{r}_k
\right\})$, we use a simple pairwise function parameterized by
first-principles calculations~\cite{Ma2008}
\begin{equation} \label{eq:exchange_func}
	J(r_{ij}) = J_0 (1 - r_{ij}/r_c)^3 \Theta (r_c - r_{ij}),
\end{equation}
where $r_{ij} = |\mathbf{r}_i - \mathbf{r}_j|$, $J_0 =
0.90490177$\;eV, $r_c = 3.75$\;\AA, and $\Theta (x)$ is the Heaviside
step function.

\subsection{Replica-exchange Wang--Landau Monte Carlo
  sampling}~\label{sec:method}

The foundation for the Wang--Landau approach is to recognize that the
canonical partition function for a system with discrete energy
levels can be written as a summation over all energies
in the form
\begin{equation} \label{eq:partition_func_wl}
	 Z = \sum_{E} g(E) e^{-\beta E},
\end{equation}
where $g(E)$ is the density of states. If $g(E)$ is known, the problem
is essentially solved since one can directly estimate the ensemble
average of any thermodynamic function of $E$ as
\begin{equation}
  \left< A(E) \right>_{_{NVT}} = \frac{ \sum_E A(E)g(E) e^{-\beta E} }{ \sum_E g(E) e^{-\beta E} }.
\end{equation}

The goal of Wang--Landau sampling is to iteratively improve the
estimate of $g(E)$ in a controlled fashion, while performing a
guided walk in energy space that eventually leads to the
accumulation of a uniform energy histogram as the estimate of $g(E)$
converges to its true value.

\subsubsection{The original Wang--Landau algorithm}

At the beginning of the Wang--Landau simulation, the desired total
energy range $E \in [E_{\textrm{min}}, E_{\textrm{max}}]$ for which
$g(E)$ should be obtained is determined.  For systems with continuous
energy domains, the total energy range is divided into bins with size
$\delta E$ appropriately chosen according the desired level of
resolution in $g(E)$.  Since $g(E)$ is unknown in the beginning of the
simulation, an initial guess of $g(E) = 1$ is assigned for all
energies.  Then, starting from an
arbitrary initial state of the system, a random walk in the
configurational space is performed by sequentially generating trial
states.  During each MC step, a new trial state $\mathbf{x}_n$ is
generated by applying an MC trial move on the current state
$\mathbf{x}_m$.  The new state is accepted according to the
probability
\begin{equation}
  P(\mathbf{x}_m \to \mathbf{x}_n) = \min{\left[ 1, \frac{g(E(\mathbf{x}_m))}{g(E(\mathbf{x}_n))} \right]}.
\end{equation}
If the trial state $\mathbf{x}_n$ is accepted, the density of states
entry for $E(\mathbf{x}_n)$ is updated as $g(E(\mathbf{x}_n)) \to
g(E(\mathbf{x}_n)) \times f$, where $f$ is the ``modification factor''
which we initially set to $f_0 = e^1$.  If the trial state is
rejected, the entry for the old state is updated as
$g(E(\mathbf{x}_m)) \to g(E(\mathbf{x}_m)) \times f$.

The random walk is continued until all energy bins have been visited
sufficiently often. Different ways of checking this condition have
been proposed~\cite{wl1,wl2,zhou05pre,belardinelli07pre}. In the
conventional version, one could maintain a histogram $H(E)$ of the
visited energies.  When all the entries in the histogram are greater
than a certain percentage of the average histogram value, the
histogram is considered to be ``flat''.  At this point, the
modification factor is reduced, for example by $f \to \sqrt{f}$, the
histogram is reset to zero, and another iteration of the random walk
is initiated.  This process is repeated until the modification factor
$f$ reaches a predefined terminal value, say $f_\text{final} = e^{1
  \times 10^{-8}}$.

\subsubsection{Replica exchange framework for massively parallel Wang--Landau sampling}

\begin{figure}
\centering
 \includegraphics[width=0.85\columnwidth]{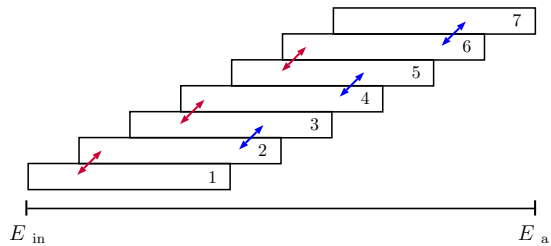}
 \caption{Partitioning the global energy range into seven windows with
   overlap $o = 75\%$.  The arrows indicate the communication pathways
   between neighboring windows for the replica-exchange attempts.  }
    \label{fig:rewl}
\end{figure}

In REWL sampling, the global energy range
$[E_{\textrm{min}}, E_{\textrm{max}}]$ is divided into $h$ smaller
windows, each of which overlaps with its nearest neighbors on both
sides with an overlap ratio $o$ (a schematic diagram is shown in
Fig.~\ref{fig:rewl}).  In each window, $m$ random walkers are
employed.  Each walker has its own $g_i(E)$ and $H_i(E)$, $0<i\leq (h
\times m)$, which are updated independently.  Once all walkers within
an energy window have individually satisfied the flatness criterion,
their estimates for $g(E)$ are averaged out and distributed among each
other before simultaneously proceeding to the next iteration.  The
simulation is terminated when the modification factors for all windows
have reached the terminal value $f_\text{final}$.

During the simulation, after every $n$ MC steps, replica exchanges
between walkers in adjacent energy windows are proposed.  For every
walker $i$, a ``swap partner'' $j$ is chosen randomly from one of the
adjacent windows.  If $\mathbf{x}$ and $\mathbf{y}$ are the current
configurations of the walkers $i$ and $j$, the two configurations are
interchanged according to the probability
\begin{equation}
	P_{_{\textrm{RE}}} = \min{\left[ 1, \frac{ g_i(E(\mathbf{x})) g_j(E(\mathbf{y})) }{ g_i(E(\mathbf{y})) g_j(E(\mathbf{x})) } \right]},
\end{equation}
where $g_i(E(\mathbf{x}))$ is the current estimate for the density of
states of the walker $i$ with energy $E(\mathbf{x})$.

At the end of the simulation, the parallel Wang--Landau method
provides multiple, overlapping fragments of $g(E)$.  These fragments
are joined at points where the slopes of $\ln g(E)$ (i.e.~$d \ln g(E)
/ dE$, the inverse microcanonical temperature) best coincide.  This
practice reduces the introduction of artificial kinks in the combined
$g(E)$ due to the joining process and minimizes artificial errors in
thermodynamic quantities~\cite{rewl2}. Any residual systematic error
is almost always less than the remaining (small) statistical error.

\subsection{Monte Carlo trial moves for coupled spin--lattice systems}

For coupled spin--lattice systems, the configu\-ra\-tional space that
one seeks to sample via MC methods consists of $2 N$ phase variables:
$ \{\mathbf{x}\} =\allowbreak \{\mathbf{r}_1,\allowbreak \mathbf{r}_2,
\cdots, \mathbf{r}_N,\allowbreak \mathbf{e}_1,\allowbreak
\mathbf{e}_2,\allowbreak \cdots, \mathbf{e}_N\}$.  For effectively
sampling this configurational space with respect to both the atomic
coordinates and the spins, we employ the following two trial moves.
\begin{enumerate}
\item \textit{Single atom displacement move} \\
  Displace the chosen atom $i$ to a random position $\mathbf{r}_i'$
  within a sphere
  centered at its original position $\mathbf{r}_i$: \\
  $\mathbf{r}_i' = \mathbf{r}_i + \mathbf{R}$, where $|\mathbf{R}| <
  R_\text{max}$
\item \textit{Single spin rotation move} \\
      Assign a new random direction to the spin of the chosen atom $i$.
\end{enumerate}
During each MC step, we randomly choose an atom and perform one of the
above trial moves at random with equal probability.  Completion of $2
N$ such MC steps constitutes a single ``MC sweep''.

\section{Results}~\label{sec:results}

Our simulations were performed on a cubic cell of size $L=20$ ($16000$
atoms; 2 atoms per unit cell) with periodic boundary conditions.  To
explore the sensitivity of the results to the particular choice of EAM
potential, we performed simulations using both Dudarev--Derlet (DD)
and the Finnis--Sinclair (FS) potentials~\cite{Dudarev2005,
  Derlet2007,Finnis1984, Finnis1986}.  The corresponding global energy
ranges were chosen to be [$-67200$\;eV, $-63200$\;eV] and
[$-67200$\;eV, $-62080$\;eV], respectively, for the DD and FS
potentials.  For both cases, $189$ energy windows with an overlap
$o=75\%$ were used, and a single walker per window ($m=1$) was
employed.  To discretize the energy space, each window was divided
into $2000$ energy bins.  Replica exchanges between neighboring
windows were proposed every $60$ MC sweeps.  With these simulation
parameters, we observed acceptance rates for the replica exchanges in
the range of $49-55\%$.  For checking the convergence of $g(E)$, an
$80 \%$ flatness criterion and a final modification factor of
$\text{ln}\,f_{\text{final}} = 1 \times 10^{-8}$ were used.  For both
potentials, the full convergence of $g(E)$ was achieved in about $1
\times 10^{8}$ MC sweeps, which took less than a week on a
$128$GB-RAM AMD Opteron cluster with InfiniBand \hbox{connectivity}.

\begin{figure}
\centering
  \includegraphics[width=\columnwidth]{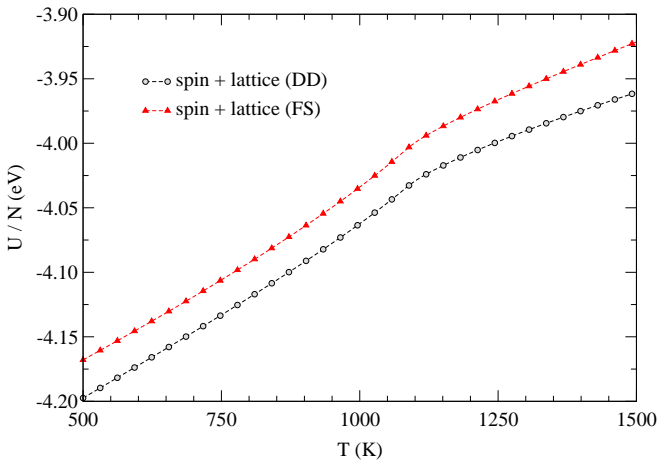}  
  \caption{ Comparison of the temperature dependence of the internal
    energy for coupled spin--lattice systems of size $L=20$ using the
    Dudarev-Derlet [``spin + lattice (DD)''] and Finnis-Sinclair
    [``spin + lattice (FS)''] potentials.  Error bars are smaller than
    the symbols.  }
 \label{fig:avgE_spin_lattice}
\end{figure}

\begin{figure}
\centering
  \includegraphics[width=\columnwidth]{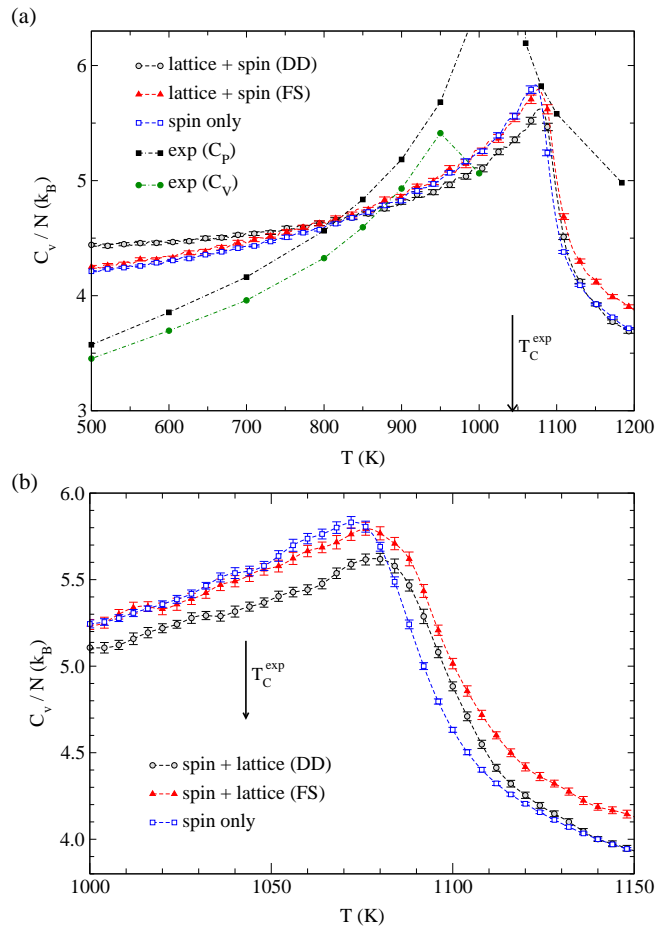}
  \caption{Specific heat as a function of temperature for $L=20$ with
    [``spin + lattice (DD)'' and ``spin + lattice (FS)''] and without
    [``spin only''] the influence of the lattice vibrations; (a)
    expanded temperature range ${[500\text{\,K}, 1200\text{\,K}]}$
    including the experimental results for $C_P$ obtained from
    Ref.~\cite{expDataCv} and the corresponding $C_V$ values
    calculated from the $C_P$ data; (b) a close-up view in the
    vicinity of the peak positions.  The vertical arrows in both (a)
    and (b) mark the Curie temperature $T_C^\text{exp} \approx
    1043$\,K as predicted by the peak position of the experimental
    $C_P$ curve.  }
 \label{fig:Cv_spin_lattice}
\end{figure}

To reduce statistical fluctuations in the estimated thermodynamic
quantities, we averaged over the results of $15$ independent runs for
the DD potential, and $11$ runs for the FS potential.
Fig.~\ref{fig:avgE_spin_lattice} shows the comparison of the
temperature dependence of the internal energy per atom obtained for
the two potentials.  For the whole temperature range considered, the
internal energy per atom obtained for the FS potential is
approximately $0.03$--$0.04$\;eV higher than that for the DD
potential.  Fig.~\ref{fig:Cv_spin_lattice} shows the specific heat
curves for the two potentials, along with the results obtained from
rigid lattice (spin only) simulations in which the atoms were held
fixed at perfect bcc lattice positions~\cite{rewl5}. Also shown in the
subset (a) are the experimental results for the constant-pressure heat
capacity $C_P$, and the corresponding $C_V$ values calculated from the
$C_P$ data~\cite{expDataCv} using the relation $C_V = C_P -
VT\alpha^2/\beta_T$, where $\alpha$ and $\beta_T$ are the thermal
expansion coefficient and the isothermal compressibility,
respectively.  Due to the lack of thermal expansion coefficient data,
$C_V$ values above $1000$\,K are not given~\cite{expDataCv}.  For a
fair comparison with the experimental results, we have added
$\frac{3}{2} k_B$ to the DD and FS results to include the contribution
of the kinetic energy based on the equipartition theorem.  For the
rigid lattice results, $3 k_B$ was added to include the contribution
of both the kinetic energy and the lattice potential energy.  The
vertical arrows in both (a) and (b) mark the Curie temperature
$T_C^\text{exp} \approx 1043$\,K as predicted by the peak position of
the experimental $C_P$ curve.  The difference between the results for
the two different embedded atom potentials is clearly larger than the
respective error bars, but both sets of results differ markedly from
the estimated values of $C_V$ extracted from experiment.  The peak in
the specific heat corresponding to the rigid lattice simulations is
approximately $30$\;K higher than the experimental Curie temperature.
The introduction of lattice vibrations further pushes the peak
position to higher temperatures by several degrees. Moreover, lattice
vibrations reduces the amplitude of the peak, an effect which is more
pronounced for the case of the DD potential.  Specific heat data for
the simple cubic Heisenberg ferromagnet~\cite{Peczak1991} has shown
that the location of the specific heat peak increases as $\sim 0.7
L^{-1/0.7}$. Hence, extrapolation of our data to infinite size would
change the result very little, as also indicated by exemplary
simulations at other system sizes.

\section{Summary}

In conclusion, we have performed highly parallel replica-exchange
Wang--Landau simulations to investigate the magnetic phase transition
in a coupled spin--lattice model parameterized for bcc iron. The high
level of precision achieved in our simulations has allowed us to make
careful comparisons between the results obtained for two different
interatomic potentials (FS and DD), and simulations performed on rigid
lattices. Such a comprehensive analysis was only possible due to the
significant speedup rendered by the parallel, replica-exchange scheme,
without any loss of accuracy or precision.  While the complete
analysis presented in this paper would take of the order of $100$
years for the serial Wang--Landau method performed on a single core
processor, we obtained all the results within a few months using the
parallel scheme.

Our results indicate that the presence of lattice vibrations only
marginally effects the transition temperature and the amplitude of the
peak in the specific heat curve. This suggests that the classical
Heisenberg model already provides a reasonable depiction of the
magnetic phase transition in bcc iron. We also find that the results
are sensitive to the particular choice of the interatomic potential,
particularly at temperatures further away from the critical
temperature $T_c$. As the temperature increases beyond $T_c$, the
specific heat obtained using the FS potential gradually deviates from
that of the rigid lattice simulations, whereas below $T_c$, a
reasonable agreement with the rigid lattice results can be observed.
In contrast, the specific heat obtained using the DD potential is
higher than that of the rigid lattice simulations up to about
$T=800$\;K, then remains smaller in comparison to the rigid lattice
results until about $T=1100$\;K, and thereafter starts to gradually
converge with the rigid lattice results.  The differences in the
results for the two EAM potentials can be attributed to the subtle
differences in the ways which the anharmonic effects are captured in
these potentials which, in turn, effect the magnetic properties of the
system via spin-lattice coupling.

\begin{acknowledgments}
  We sincerely thank Ying Wai Li and Markus Eisenbach for informative
  discussions.  
This work was sponsored by the ``Center for Defect Physics'', an Energy Frontier Research Center of the Office of Basic Energy Sciences, U.S. Department of Energy.  
We also acknowledge the computational
  resources provided by the Georgia Advanced Computing Resource
  Center, a partnership between the University of Georgia's Office of
  the Vice President for Research and Office of the Vice President for
  Information Technology.
\end{acknowledgments}

\end{document}